\documentstyle[12pt]{article}
\def\dsp{\displaystyle}

\def\dddot#1{\mathinner{\buildrel\vbox{\kern5pt\hbox{...}}\over{#1}}}
\def\n{\nonumber}
\def\o{\omega}
\def\ha{\mbox{$\frac{1}{2}$}}
\def\e{{\rm e}}
\def\ve{{\varepsilon}}

\def\d{{\rm d}}

\def\b{\beta}
\def\L{\Lambda}
\def\O{\Omega}
\def\be{\begin{equation}}
\def\ee{\end{equation}}
\def\bq{\begin{eqnarray}}
\def\eq{\end{eqnarray}}
\def\beq{\begin{eqnarray*}}
\def\eeq{\end{eqnarray*}}
\def\p{\partial}

\def\({\left(}
\def\){\right)}

\begin{document} \openup6pt
\titlepage
\vspace*{2cm}
\begin{center}
{\Large\bf Adiabatic Invariants and  Mixmaster Catastrophes}

\vspace{1cm}

{\large S. Cotsakis, R.L. Lemmer$^1$ and  P.G.L. Leach$^{1,2}$}
\vspace{5mm}

Department of Mathematics\\ University of the Aegean\\ Karlovassi 83 200,
Samos\\
Greece \\
(skot@aegean.gr, lemmer@ph.und.ac.za, leach@aegean.gr)
\end{center}
\vspace{5mm}

$^1$ Permanent address:  Department of Mathematics and Applied Mathematics,
University of Natal, Private Bag X10, Dalbridge 4014, Durban 4001, South Africa
\vspace{5mm}

$^2$ Member of the Centre for Theoretical and Computational Chemistry,
University of Natal, Durban, and associate member of the Centre for Nonlinear
Studies, University of the Witwatersrand, Johannesburg, South Africa
\vspace{5mm}


\newpage
\titlepage
\begin{center}
\section*{Abstract}
\end{center}
 We present a rigorous analysis of the role and uses of the adiabatic invariant
in the Mixmaster dynamical system. We propose a new invariant for the global
dynamics which in some respects has an improved behaviour over the commonly
used one. We illustrate its behaviour in a number of numerical results.
We also present a new formulation of the dynamics via
Catastrophe Theory. We  find that the change from one era to the next
corresponds to a fold 
catastrophe, during the Kasner shifts the potential is an Implicit Function
Form whereas, as the anisotropy dissipates, the Mixmaster potential must
become a Morse 0--saddle. We compare and contrast our results to many known
works on the Mixmaster problem and indicate how extensions could be achieved.
Further exploitation of this formulation may lead to
a clearer understanding of the global Mixmaster dynamics.

\renewcommand{\thesection}{\Roman{section}}

\newpage
\section {Introduction}
During the past eight years or so interest in the Mixmaster universe
\cite{hobill} has increased dramatically. There are at least two reasons for
this noticeable increase. Firstly, there was already a considerable amount of
background work concerning the basic dynamical issues of the model
\cite{BKL,misner,bar,bogo} and it is probably fair to say that this model
(diagonal Bianchi IX) was studied more than any other homogeneous cosmology
(with a probable exception of inflationary issues). The picture drawn from that boby of
work was already quite rich (the actual system was also integrated numerically
by Zardecki \cite{z} but some of his results later criticized
as incompatible with those of BKL  in \cite{Ma,FM}) and allowed for further
generalizations  to be considered.

Examples of this kind of generalized problems included
Kaluza-Klein extensions \cite{kaluza} and the search for complicated Mixmaster
behaviour in other theories of gravitation \cite{hog}. All these
generalizations were motivated mainly by an idea and results first obtained by
 J.D. Barrow \cite{bar} almost another decade earlier,
that the well--known  BKL-Misner oscillatory,
Mixmaster behaviour should in fact be viewed as
an example of the manifestation of chaotic (erratic, unpredictable) structures
in general relativity. It was therefore natural to examine how common such a
behaviour could be. (It was also known \cite{BK}  that departures from such an
evolutionary scheme could be obtained with the addition of scalar fields).
In fact, many of those results were quite interesting and contributed much to a
sustained interest in the Mixmaster dynamics during the eighties.

Interest in the Mixmaster universe suddenly peaked with the appearence of the
first works on numerical experiments in 1989-90 \cite{num1}. Those results (and
others which followed  \cite{num2}) were conflicting in the sense that
 in most cases the standard picture \cite{bar} was challenged to the effect
that many workers in the field felt that a re-examination of the original
conclusions concerning the existence of chaos in the Mixmaster approach to the
initial singularity was necessary. At the same time other works appeared which
either criticized
\cite{pullin} the use of
some essentially coordinate dependent measures of chaoticity or indicated
\cite{SB,DYD} that
some sort of chaotic behaviour should be present in the Mixmaster dynamics.

It was felt necessary \cite{cot1} that perhaps
an analytical approach, known as  {\em Painleve analysis,}
which did not share the `defects' of numerical work could lead to
more reliable results concerning the existence or non-existence of chaotic
behaviour in the Mixmaster dynamics. Initial results \cite{co-co} pointed to
the direction of integrability whereas later it was realised \cite{con-con}
that the situation was more complex. It is now understood that this analysis
cannot be used to obtain reliable results concerning the question of
chaoticity in this model.

Very recently Cornish and Levin \cite{cor-lev} using fractal methods resolved
the long--standing debate concerning the issue of chaoticity in the Mixmaster
universe, showing that the system is indeed chaotic. Their analysis not only
confirms the earlier ergodic results of Barrow \cite{bar} but quantifies the
chaotic behaviour of the model by calculating a special set of numbers
(topological entropy, multifractal dimension etc) relevant to the true
dynamics. It therefore appears that all ambiquities concerning this issue have
now disappeared.

It is perhaps encouraging
 that several issues about the Mixmaster dynamics {\em
not connected to the question of chaos} have been studied by several authors.
It is indeed true that many problems in the dynamics of this model still remain.  First of
all it is still unknown how local the BKL analysis is \cite{bar-BKL}. Another
problem is borne out of earlier results of Moncrief \cite{mon} and is related
to the role that  the Geroch transformation plays for the true dymanics. Still
another issue has to do with the description of the Mixmaster universe from the
point of view of dynamical systems theory \cite{rendal}. We therefore believe
that the rich Mixmaster behaviour will continue to attract  the interest of
cosmologists for  some time to come.

{\em FIGURE 1.  TO BE INSERTED HERE}

The purpose of this paper is to provide a rigorous analysis of the role and
uses of adiabatic invariants in the Mixmaster problem by carefully examining
the adiabatic invariant commonly used  and introducing a new and in a sense
improved invariant for the global Mixmaster dynamics. We then  reformulate
 the main characteristic of the dynamics via a new, simpler
 technique along the lines of Catastrophe theory. We believe that further
exploitation of this formulation may help to unravel certain global dynamical
properties of the Mixmaster universe.

The plan of this paper is as follows. In Section II we introduce
our new adiabatic
invariant for the Mixmaster system which behaves better than the standard one
and is explicitly time--independent in the appropriate coordinates.
We also perform a
numerical simulation of the corner--run part of the evolution and we give an
interpretation of the results using Misner's Hamiltonian picture. In Section
III,
we apply Catastrophe Theory as a means to gain a better understanding of the
complicated behaviour of the model. The main result of this Section  is that the
passage from one era to the next corresponds, in the language of Catastrophe
Theory, to a fold catastrophe which in turn may provide a potentially
new way to view the
global evolution. In the last Section, we   compare  our results to previous
work and point out how generalizations to higher dimensions could be
obtained.
\section{Adiabatic Analysis}
Subject to a
couple of overall approximations the motion of the `universe point' $\b =(\b_+
,\b_- )$ is governed by the Mixmaster Lagrangian
\be
L_{full} = \ha\L^{\ha}\left(\b_+'{}\!\!^2 + \b_-'{}\!\!^2\right) - 2\L^{-\ha}\e^{-4\O}V(\b),
\label{1.1}
\ee
where $\b_+$ and $\b_-$ are related to the shape parameters of the model, the
volume parameter, $\O$ plays the r\^{o}le of time, primes denote
differentiation with respect to $\O$ and  $\L$ is a function
of $\O$ which evolves according to
\be
\L ' = -4\e^{-4\O}V(\b).   \label{1.2}
\ee
Here $V(\b)$ is the standard, curvature anisotropy  (Mixmaster)
potential given by
\be
V(\b) = \frac{1}{3}\e^{-4\b_+} - \frac{4}{3}\e^{-\b_+}\cosh (\sqrt{3}\b_-)
+ \frac{2}{3}\e^{2\b+}[\cosh (2\sqrt{3}\b_-) - 1] + 1. \label{1.3}
\ee
There is also the energylike equation
\be
4 = \b_+'{}\!\!^2 + \b_-'{}\!\!^2 + 4 \L^{-1} \e^{-4 \O} V(\b). \label{1.4}
\ee
We begin our analysis by re-examining the role and consequences of the
adiabatic invariant used implicitly (or explicitly) in most Mixmaster
calculations.
According to Misner \cite{misner}
the most important asymptotic form of the Mixmaster potential is when
$\mid\b_-\mid$ is small and slowly varying and
$\b_+\longrightarrow +\infty$ , that is, $V(\b )\sim1+ 4\b_-^{2} e^{2\b_+}.$
In this case  $L_{full}$ can be considered as the Lagrangian for the $\b_-$
motion; that is, it can be supposed approximately that there exists a system
with one degree of freedom, $\b_+$, depending on the slowly varying parameter
$\b_-$. This reduced system is described by the lagrangian,
\be
L_{reduced} = \ha\L^{\ha}\b_-'{}\!\!^2 - 8\L^{-\ha}\e^{2\b_+-4\O}\b_-^2 \label{1.6}
\ee
in which a term $O(\e^{-4\O})$ has been neglected and $\L$ and $\b_+$ are again
 functions of $\O$ through the solutions of (\ref{1.2}) and
(\ref{1.4}) with the potential as modified.  The Lagrangian (\ref{1.6}) is that
of a time--dependent oscillator of slowly varying frequency given by
\be
\o_- = 4\L^{-\ha}\e^{\b_+-2\O}. \label{1.7}
\ee
Then, for the reduced system, there exists the adiabatic invariant
\bq
\Sigma &=& \frac{E_-}{\o_-}\n\\
&=& \frac{1}{8}\L \e^{2\O-\b_+}(\b_-'{}\!\!^2 + \o_-^2\b_-^2), \label{1.8}
\eq
where $E_-$ are the energy level sets and $2\pi\Sigma$ represents the area of
appropriate domains bounded by curves passing through points in (the two
dimensional phase) space $(\pi_- ,\b_- )$.
It can be shown, by adapting the methods of \cite{A1},
 that this is also an adiabatic invariant of the `full'
Mixmaster system (two degrees of freedom) with $\b_-$ slowly varying (but not
necessarily small).
We stress this point since it is important to remember that the motion of the
universe point described by $L_{reduced}$ is only approximately true and the
true dynamics in this case should be thought of as
 that given by $L_{full}$ with $\b_-$ slowly varying.

The adiabatic invariant (\ref{1.8}) (considered originally by Misner
\cite{misner})
 is exactly the one
proposed by Lorentz at the first Solvay Congress in 1911 \cite{L3,WE}.
A precise
mathematical statement concerning its range of validity was given by
 Littlewood \cite{L2}, in the sixties.  Littlewood showed that (in
our notation)
\begin{description}
\item {\it (i)} $\Sigma = c+O(\ve)$;
\item {\it (ii)} $\bar{\Sigma} = c
+O(\ve^2)$ ($\bar{\Sigma}$ is the average of $\Sigma$ over the local period
$2\pi/\o_-$.)
\item {\it (iii)} there is no improvement over {\it (i)} or {\it (ii)}
\item {\it (iv)} $\Sigma(\infty)-\Sigma(-\infty) = O(\ve^n)$ for some
specific $n$
\end{description}
provided $\o_-$ satisfied certain assumptions which give a measure,
$\ve$, to the expression `slowly varying'.  These assumptions are
\bq
\o_->b_0,&&\o_-^{(n)} = \frac{\d^n\o_-}{\d t^n}\rightarrow 0\quad\mbox{as}\quad
t\rightarrow\pm\infty\quad (n \geq 1)\n\\
|\o_-^{(n)}|<b_n\ve^n (n\geq 1),&&\int_{-\infty}^{\infty}|\o_-^{(n)}|\d t <
b'_{n-1}\ve^{n-1},\eq
where the $b$s are constants.
  A consequence is that, since
\[
|\o_- (\tau')-\o_- (\tau)|\leq\int_{\tau}^{\tau'}|\dot{\o_-}|\d t \rightarrow 0
\]
as $\tau$, $\tau' \rightarrow -\infty$, $+\infty$ respectively, we have
\[\o_- (-\infty) = \o_- (+\infty).\]
Strictly we need not go as far as $\pm\infty$, but the in--channel time is
supposed to be long.

Following Arnol'd \cite{A} a
function $I(q, p; \ve t)$ is an adiabatic invariant of a Hamiltonian system
\be
\dot{p} = - \frac{\p H}{\p q} \qquad \dot{q} = \frac{\p H}{\p p} \qquad H =
H(q, p; \ve t) \label{2.1}
\ee
if $\forall \kappa > 0\, \exists\, \ve_0 $ such that if $\ve < \ve_0$ and $0 < t <
1/\ve_0$,
\be
|I(q(t), p(t); \ve t) - I(q(0), p(0); 0)| < \kappa. \label{2.2}
\ee
The action variable of the corresponding autonomous problem is always an
adiabatic invariant and it is for this reason that (\ref{1.8}) can be selected
as the adiabatic invariant.

However, the perpetuality (cf. \cite{A1}) must be calculated and verified.
Previous works
provide no information about the behaviour of $\o_-$ and certainly no measure of
$\ve$.  Fortunately it is a fairly straightforward matter to test the validity
of the use of the adiabatic invariant (\ref{1.8}).  We can simply
integrate the equations of motion numerically and substitute the numbers into
(\ref{1.8}) to observe the variation of $\Sigma$ with $\Omega$.

In the channel r\'{e}gime the equations governing the motion of
the system point are
\bq
\L ' &=& - 16\e^{2\b_+ - 4\O}\b_-^2 \label{2.3}\\
\b_+ ' &=& \pm\left( 4-\left(\b_- '\right)^2 -
\frac{16}{\L}\e^{2\b_+-4\O}\b_-^2\right)^{1/2} \label{2.4}\\
\b_- ''&-&
\frac{8}{\L}\e^{2\b_+-4\O}\b_-^2\b_- ' +
\frac{16}{\L}\e^{2\b_+-4\O}\b_- = 0.   \label{2.5}
\eq
The positive sign in (\ref{2.4}) applies until the expression under the
square root sign
becomes zero.  Equations (\ref{2.3})--(\ref{2.5}) are those of Misner
\cite{misner}
except that $\b_0$ has not been introduced.  We take initial conditions
consistent with the assumptions governing motion in the channel. 
We integrate the system (\ref{2.3}--\ref{2.5}) numerically using
an implementation of the Runge-Kutta scheme \cite{lc95}. The results are
illustrated in Figure 2.  It is quite evident that these computations do not
support the use of the adiabatic invariant over the whole time spent inside
the channel.

{\em  FIGURE 2. TO BE INSERTED HERE.}

Let us now consider some consequences of the above analysis.
Upon the introduction of a relative coordinate
\be
\b_0 = \b_+ - 2\O  \label{1.9}
\ee
and, with the help of eqs (\ref{1.7}) and (\ref{1.8}),
eq (\ref{1.4}) becomes
\be
0 = \b_0'{}\!^2 + 4\b'_0 + \frac{\Sigma\o_-^2}{2}\e^{-\b_0}. \label{1.10}
\ee
On the assumption that $\b'_0$ is small, $\b_0'{}\!^2$ is neglected and
(\ref{1.10}) gives immediately
\be
\b_0 = \log(\O_0 - \O) + const \label{1.11}
\ee
which leads to $\b_0 \longrightarrow -\infty$ as $\O$ approaches the critical
value $\O{_0}$.  This means \cite{misner} that the particle leaves the channel and returns bouncing in the
triangular region.

Consider now the  assumption that is usually made namely,
that $\b'{}^2_0$ may be neglected. We claim that this assumption
leads to a solution which is only asymptotically correct.
To see this, notice that it  follows  easily
 that the solution of (\ref{1.10}) is given implicitly by
\be
\O_0 - \O = \frac{1}{4}\left\{a^{-1}\e^{\ha\b_0} + \sqrt{a^{-2}\e^{\b_0}
- 1}\right\}
- \ha\log\left\{a^{-1}\e^{\ha\b_0} + \sqrt{a^{-2}\e^{\b_0} -
1}\right\}, \label{1.12}
\ee
where
\be
a^2 = \frac{\o^2_-\Sigma}{8}, \label{1.13}
\ee
is assumed constant.  Misner's original solution (\ref{1.11}) is, in a
sense, asymptotic 
to (\ref{1.12}) and our assertion follows.

Of more interest, however, is the actual equation for
$\b'_0$, which follows immediately from (\ref{1.10}),  {\em viz.}
\be
\b'_0 = -2 \pm 2\sqrt{1 - a^2\e^{-\b_0}}. \label{1.14}
\ee
(The upper sign applies for the initial motion in the channel since $|\b'_+| <
 2.$)  Eq. (\ref{1.14}) is valid only if
\be
\e^{\b_0} \geq a^2 .\label{1.15}
\ee
Inserting this into eq (\ref{1.12}) we find  that
\be
 \O \leq \O_0 -\frac{1}{4}\hspace{1cm}or\hspace{1.5cm}\b_0\geq -0.60
+const. \label{1.16} 
\ee
In other words $\O$ never reaches the critical value $\O_0$ and $\b_0$ cannot
go  to $-\infty$ which is what is necessary
for the particle to return to the triangle from the channel.
 (Of course, in practice, it is only sufficient to have
$\b_0$ large and negative for the particle to resume bouncing with the wall
in the triangular box as soon as it leaves the channel, but even for this to
happen the bound in Eq. (\ref{1.16}) seems too stringent.)

We are now ready to introduce a new invariant which in many ways behaves better
than $\Sigma$.
It is a well--known fact that the time--dependent oscillator described by the
Hamiltonian
\be
H = \ha(p^2 + \o^2(t)q^2) \label{3.1}
\ee
possesses the first integral
\be
I = \ha\left((\rho p - \dot{\rho}q)^2 + \left(\frac{q}{\rho}\right)^2\right),
\label{3.2}
\ee
which is known as the  Ermakov--Lewis invariant \cite{EM}, provided that the
auxiliary variable is a solution of
\be
\ddot{\rho} + \o^2(t)\rho = \frac{1}{\rho^3}. \label{3.3}
\ee
Furthermore the solution of (\ref{3.3}) has been given by Pinney \cite{P} in
terms of the linearly independent solutions of
\be
\ddot{v} + \o^2(t)v = 0. \label{3.4}
\ee

The Lagrangian (\ref{1.6}) gives directly the Hamiltonian
\be
H_{reduced} = \ha\L^{-\ha}\Pi_-^2 + 8\L^{-\ha}\e^{2\b_+-4\O}\b_-^2, \label{3.5}
\ee
where
\be
\Pi_- = \L^{\ha}\b_-'. \label{3.6}
\ee
The reduced Hamiltonian (\ref{3.5}) is not precisely of the form of (\ref{3.1}) since
the coefficient of $\Pi_-^2$ is not constant and we have the equivalent of an
harmonic oscillator of variable mass.  However, following the procedure
detailed by Leach \cite{L83}, for the treatment of a Hamiltonian of the form of
(\ref{3.5}) we introduce a change of time scale
\be
T = \int \L^{-\ha}(\O)\d\O \label{3.7}
\ee
so that the Mixmaster system in this regime is described by
\be
\tilde{H}_{reduced} = \ha\Pi_-^2 + \ha\o^2(T)\b_-^2 \label{3.8}
\ee
in which $T$ is now the independent variable and
\be
\o^2(T(\O)) = 16\e^{2\b_+(\O)-4\O}. \label{3.9}
\ee
Under the generalised canonical transformation \cite{L82}
\be
Q = \frac{\b_-}{\rho} \qquad P = \rho p - \dot{\rho}q \qquad \tau =
\int\rho^{-2}(T)\d T \label{3.10}
\ee
(\ref{3.8}) is transformed to
\be
\bar{H} = \ha(P^2 + Q^2)
\ee
provided that $\rho(T)$ is a solution of (\ref{3.3}) (with $\o^2(T)$ from
(\ref{3.9}) instead of the $\o^2(t)$).  $\bar{H}$ is the Ermakov--Lewis
invariant in the canonical variables in which it becomes free of explicit
dependence on time.

It is $\bar{H}$ which should be used instead of the adiabatic invariant used by
Misner.  However, there is a problem.  In the new variables the evolution of
the oscillator is easily described, but that of $\b_+$ becomes unmaneageable as
the Ermakov--Lewis invariant does not lead to a significant simplification of
(\ref{2.4}).  We must resort to numerical computation and this may as well be
performed in the original variables.

{\em FIGURES 3,4,5. TO BE INSERTED HERE}

 Over the interval $0<\omega < 30 000$, $\beta_+$
increases essentially linearly with $\omega$.
There is no indication that it approaches $-\infty$ and as the deviation
from strict linearity is so small, it can only be expected to take a long time
to approach zero. In the meantime the potential well proceeds outwards as is
evident from Figure 4 at which the contour is that corresponding to the energy
of the motion. We see that the wall outmarches the $\beta_+$ value of the
universe point. Over this period, as depicted in Figure 5, the amplitude of the
$\beta_-$ motion increases in a strictly monotonic fashion with a rate of
increase increasing with time.

It is clear from these results that, as the particle
moves along the channel, the potential walls are receding.  Initially this does
not present a problem as the $\b_+$ velocity of the particle is sufficiently
large. 
However, there is a critical time $\O_0$ when the walls `leave the particle
behind'.  Hence the particle finds itself in the triangular region again and no
longer in the channel.
Note that, as $\b_+$ is never less than zero, the particle does not `turn
around' in the sense that is sometimes described.
\section{A Catastrophe Description}

In what follows we present  a novel way to  describe the qualitative
differences  of the  main stages
in the evolution of the Mixmaster
universe. As we shall see, the standard interpretation can be reached quite
naturally and independently by this line of thought. This approach is
established by the application of Singularity Theory \cite{Lu76} and in
particular by that branch of singularity theory known as Catastrophe Theory
\cite{Lu-Gil}. Catastrophe theory studies changes in the equilibria of
potentials as the 
control parameters  of the system change. The local properties of the
potential in a gradient or a dynamical system are determined by a sequence
of theorems  such as the Implicit Function Theorem of
advanced calculus, the Morse Lemma \cite{Morse} and the Thom Theorem \cite{Thom}.

We consider the Mixmaster universe as a gradient system described
by the potential (\ref{1.3}). We choose as a control parameter the volume
(time) parameter $\Omega$. We  set  $\nabla
=(\p/\p\b_+, \p/\p\b_-)$ and denote the Hessian matrix by
\be 
V_{ij} =
\(\begin{array}{cc}
\frac{\dsp{\partial ^2 V}}{\dsp{\partial \beta _{+}^2}} & \frac{\dsp{\partial
^2 V}}{\dsp{\partial
\beta _{+}\partial \beta _{-}}} \\ \frac{\dsp{\partial ^2 V}}{\dsp{\partial \beta
_{+}\partial \beta _{-}}} & \frac{\dsp{\partial ^2 V}}{\dsp{\partial
\beta _{-}^2}}
\end{array}\). \label{hess}
\ee
 Then, for the Kasner--to--Kasner evolution described by the potential
\be
\b_+ \longrightarrow -\infty ,\hspace{1.5cm}V(\b )\sim\frac{1}{3}e^{-8\b_+} ,
\label{kasner}
 \ee
 we find
\be
\nabla V=\(-\frac{8}{3}e^{-8\b_+},0\)\neq 0. \label{IFF}
\ee
This means that during an era the Implicit Function Theorem applies and there
are no critical points. (This also implies that there is a smooth change of
coordinates which makes the potential (\ref{1.3})  depend on only one of the
variables, say $\b_+$. We see that the form of the potential (\ref{kasner})
can be deduced from the general form (\ref{1.3}) by using
 this argument without resorting to any sort of approximations.)

Secondly we  examine the structure of the potential in the
neighborhood of the isotropy point  (0,0) given by
\be
(\b_+ , \b_- )\sim (0,0),\hspace{1.5cm}V(\b )\sim 16(\b_+^2 +\b_-^2 ),
\label{iso}
\ee
After some straightforward manipulations we find that
\be
\nabla V=0. \label{morse1}
\ee
We see that the conditions for the validity of the Implicit Function Theorem
are no longer satisfied. Equation (\ref{morse1}) implies that near the
isotropic point $(0,0)$ the Mixmaster universe is in a stable equilibrium
state. To see this we determine the stability properties of this state by
finding the eigenvalues of the Hessian matrix, $V_{ij}$. Firstly after  a
tedious calculation we find
\be
det_{(0,0)} V_{ij} =\frac{256}{3}\neq 0.
\ee
Also the eigenvalues of $V_{ij}$ are $\lambda_1 =\lambda_2 =16$. This means
that 
due to the Morse theorem \cite{Morse} there is a smooth change of variables so
that the potential in this case takes the form
\be
V=M_0^2 =\lambda_1^2 +\lambda_2^2 =16(\b_+^2 +\b_-^2 ) .
\ee
Not surprisingly this is exactly the form of the potential that Misner
found in this case. $M_0^2$ stands for the Morse 0-saddle which is the only
i-saddle that is stable for two--dimensional gradient systems ({\it cf}
 \cite{Lu-Gil}). Thus the point (0,0) is a Morse critical point
(isolated, nondegenerate). In particular, the potential in this case is
structurally stable.

Lastly we examine the corner--run evolution
which turns out to be the most
interesting from the point of view of Catastrophe Theory. In this
case we find
\be
\nabla V=0  \label{morse2}
\ee
and 
\be
det V_{ij} =-8172\b_-^2 e^{8\b_+}.
\ee
It is clear that for all points on the $\b_+$-axis ($\b_- =0$) we have
\be
det_{(\b_+,0)} V_{ij} = 0.
\ee
This implies that in the channel region all points which lie on the 
 $\b_+$-axis are non-Morse critical points. In this case we can cast the
potential in a canonical form by adopting a procedure known as the Thom
Splitting Lemma \cite{Gr-Mey}. We split the potential into a Morse part and a
non-Morse part according to the number of the vanishing eigenvalues of the
Hessian matrix for this case. These are found to be
\be
\lambda_1 =0,\hspace{1.5cm}\lambda_2 =32e^{4\b_+},
\ee
which due to a theorem of Thom  \cite{Thom} guarantees that there is a
smooth change of 
variables that puts the potential (in the channel) in the decomposed form
\be
V(\b )=\b_+^3 +\alpha\b_+ +32\b_-^2 ,
\ee
with $\alpha\neq 0$. The first two terms in this potential (the non-Morse part)
form what is known as 
the fold 
Catastrophe ($A_2$) and it is the simplest of the seven elementary catastrophes
first discussed by  Thom in \cite{Thom}.  The Morse part of the above
decomposition is unaffected by perturbations so it is only necessary to
study how the qualitative properties of the catastrophe function 
$A_2 =\b_{+}^{3} +\alpha\b_+$ are changed as the control
parameter changes. When  
$\alpha >0$ there are  no critical points whereas  $\alpha <0$ gives two
critical points namely, $\b_\pm =\pm\sqrt{-\alpha}$. The case $\alpha =0$ is
the separatrix in the control parameter space between functions of two
qualitatively different types (no critical points and two critical points).

Our interpretation of the above results uses the {\em delay convention } of
Catastrophe Theory (see for instance \cite{Lu-Gil}). Since, as
we have shown,  during the bounces of the 
point with the walls in the triangular box there are no critical points, we
imagine that when the point enters the channel has $\alpha <0$. Then as it
moves inside the channel the degenerate point $\alpha =0$ is reached (the
stable minimum disappears into the degenerate critical point). At this
instant $\b_-$ is no longer small, there are no critical points and the
system jumps to the lowest of the two minima (the stable attractor) of
$\alpha >0$. This produces a (point) shock wave which is the simplest
elementary catastrophe (fold). This, in turn, means that the system (point)
has found itself bouncing again inside the (now larger) triangular box.

\section{Conclusions}
Our adiabatic analysis relates to   the well--known issue of the
so--called `anomalous' behaviour discussed previously analytically by Berger  in
Ref. \cite{num2} and numerically in ref. \cite{Mo}, \cite{Berger}.
Physically, this in--channel behaviour
 appears only when the initial value of the
so-called BKL parameter $u$ is sufficiently large (a BKL `long era'). However,
the required value of $u$ becomes larger (and therefore less probable) as the
singularity is approached. This in turn means that the in--channel behaviour
becomes less probable as the Mixmaster singularity is approached.

The recent demonstration of chaoticity by Cornish and Levin  \cite{cor-lev}
via the
existence of a Mixmaster fractal strange repellor may be seen in the light of
the non--adiabatic Mixmaster evolution discussed here.
It is interesting
to point out that the problem of the existence of an adiabatic invariant for
higher dimensional generalizations of the Mixmaster universe \cite{kaluza}
or in higher derivate extensions \cite{hog} (wherein chaotic behaviour 
may be absent) is a nontrivial
one and one expect that the usual difficulties \cite{A1} present in dynamical
systems with more than two degrees of freedom  exist in this problem too.

Our numerical results parallel those given in \cite{Mo,Berger} in the
following respects: In those references,
 figures equivalent to our Fig. 5 are given but
the variables $\beta_{\pm}/\O$ are plotted there (mixing bounces) rather than our variables
$\beta_{\pm}$. We stress that no confusion must arise in this respect
since, in the former variables  the trajectory associated with a single era appears
to move outward along a corner and then inward again while, here the
motion is strictly outward. Further, as is clearly emphasized by Berger in
\cite{Berger}, the angle of the minisuperspace trajectory becomes ever closer to the
perpendicular to the ray down the corner as the era progresses towards the
singularity. A change of era occurs when the trajectory points inwards rather
than outwards with respect to  this perpenticular direction.

We hope that our reformulation of the problem in terms of Catastrophe
theory in Section III
 may be further used to examine questions of current interest such as,
for instance, issues connected with the occurence of chaotic behaviour. In some
sense, our Catastrophe results  correspond to just
 Mixmaster {\em statics}. Further dynamical issues could be
addressed if one considers the Mixmaster system as a gradient system as is
usual in Catastrophe discussions of dynamical systems.

Another issue that is raised by our formulation is the influence of changing
the time parameter on the character of the Catastrophe. It is well--known
that there exist different time parametrizations for the description of the
Mixmaster dynamics (see, for instance, \cite{Berger}). It is therefore
appropriate to ask  how the Catastrophe profile of the Mixmaster dynamics is
affected by different choices of time. Although the answer to this question is
uncertain at present, we believe that a physically relevant formulation should
be unaffected by different time choices.

S C thanks the Research Committee of the University of the Aegean for its
continuing support and the National Foundation for Research and Technology and
a grant.  P G L L and R L L thank the Foundation for Research Development of
South Africa and the University of Natal for their continuing support.  We
thank the Institute for Nuclear Research of the Academy of Sciences of the
Czech Republic and MAPMO, Universit\'{e} d'Orl\'{e}ans, for the provision of
computing facilities and  L. Cair\'{o} for the use of his Runge--Kutta code
and H. Pantazi for an independent check of the references.

\end{document}